\documentclass{article}

\usepackage[dblblindworkshop, final]{neurips_2026}
\workshoptitle{Trustworthy AI for Good (AI4GOOD)}

\usepackage[utf8]{inputenc} 
\usepackage[T1]{fontenc}    
\usepackage{hyperref}       
\usepackage{url}            
\usepackage{booktabs}       
\usepackage{amsfonts}       
\usepackage{nicefrac}       
\usepackage{microtype}      
\usepackage{xcolor}         
\usepackage{amsmath}
\usepackage{graphicx}
\usepackage{wrapfig}
\usepackage{float}
\usepackage{amsmath,amssymb}
\usepackage{amsthm}

\theoremstyle{plain}
\newtheorem{theorem}{Theorem}

\newtheorem{lemma}[theorem]{Lemma}
\newtheorem{corollary}[theorem]{Corollary}

\title{Needs More Ratings: Publishing Community Notes by Precedent}

\author{%
  Alec ~Laprevotte \\
  Harvard University\\
  \texttt{alaprevotte@g.harvard.edu} \\
}

\begin{document}

\maketitle

\begin{abstract}
Community-driven content moderation has become commonplace on platforms such as X and Meta's Threads, where regular users can add helpful context to misleading posts. Recently, LLMs have also been employed to write and submit notes faster, making note selection the main bottleneck. X's Community Notes' bridging-based algorithms publish only one in ten proposed notes and often do so hours after the original post's peak engagement, limiting the reach and corrective value of the note. Drawing from social choice, we propose a guarded precedent-closure rule, which publishes notes if their rater evidence is stronger, on multiple tracked dimensions, than the evidence of earlier published notes. In a chronological offline replay on X's public data (857,256 notes created Nov 2025--Aug 2026), the rule recovers 75.4\% of the incumbent's CRH decisions, reaches shared notes a median 0.9 hours earlier, and adds 35,945 publications. Population-sampled raters grade these additions on par with the notes only the incumbent publishes, and above those admitted at matched coverage by a reimplementation of X's core scorer with a relaxed threshold or shrinkage.
\end{abstract}

\section{Introduction}
Large public conversation platforms have often struggled with effectively limiting disinformation on their site. Fabricated content disseminates further than truthful news \citep{Vosoughi2018}, garnering vast attention in a very short time window \citep{Pfeffer2023} and complicating fact checking. Professional fact-checking often arrives only after this initial burst: during the 2022 U.S. midterms, the median institutional fact-check was published four days after a misinformation narrative first appeared \citep{Wack2026}. As such, many platforms have shifted from internal, centralized content moderation teams to crowd-sourced systems, where users themselves can counter disinformation as they see it circulate. X's Community Notes (formerly Twitter's Birdwatch) was the first system to implement this solution at scale: contributors can append their own context to misleading posts, which would then be displayed globally if deemed helpful enough by ideologically diverse users \citep{wojcik2022}. In January 2025, Meta also adopted X's open-sourced algorithm across Facebook, Instagram, and Threads \citep{kaplan_meta2025communitynotes}, and YouTube and TikTok have piloted analogous features \citep{Li_2025}. This approach has matched professional fact-checkers in accuracy with small, politically balanced crowds \citep{allen_arechar_2021, Martel2023CrowdsCE}, and reduced reshares on misleading posts by 45–62\% \citep{chuai2024, renault2024, Slaughter_2025}. By the end of 2024, nearly one million users had written or rated at least one note on X \citep{arjmandilari2025}, and these have been seen as more trustworthy than context-free misinformation flags largely because they explain why a post is misleading \citep{drolsbach2024}.

Recently, large language models (LLMs) have been introduced to scale the volume of corrective content. Contributors can submit notes with their own LLM pipelines, which are then identically rated by human users \citep{Li_2025}, lowering the barrier to writing helpful context \citep{li2026aifactcheckingwildfield}.

Despite the abundance of notes written, selection is still a bottleneck. 9\% of all notes on X are published globally (by reaching the Currently Rated Helpful (CRH) status) and appear on average 26 hours after a post is published, well after peak visibility for most misleading posts \citep{truong2025}. Meta also reports that their closed-source, community-notes based algorithm only published 6\% of proposed notes are, concluding that the current system is too slow and too vulnerable to manipulation to expand their pilot to additional countries \citep{oversightboard2026communitynotes}. Studies also found that human authors whose first note reached CRH status were far more likely to contribute again, meaning that low publication rates detrimentally affect participation \citep{arjmandilari2025}.

To this end, we propose a new guarded precedent-closure selection rule that publishes notes if the weight of the rater evidence backing them is greater than that of previously published notes. Concretely, we track 8 rating metrics for each note (support and opposition from raters on each of the two global poles of the bridging factorization, rating counts per pole, and exposure to partisan raters), and publish those with greater support and weaker opposition than $K$ previously admitted notes. By construction, this rule therefore extends bridging algorithms by only selecting notes whose evidence coordinate-wise dominates that of $K$ previously published notes, while also satisfying dominance coherence, support-addition monotonicity, and two-sided history monotonicity.

Our contributions are threefold. First, we investigate X's current note selection mechanism, identify its assumptions and catalog the effects of its design choices (shrinkage, a volume-dependent threshold, and nonmonotonicity under supporting ratings). We then propose a new publication rule, provide axiomatic guarantees, show that every admission is certified by $K$ historical acceptances it coordinate-wise dominates, and derive a certified opposition margin, where  no $m$ added ratings can revoke an admission's base certificate under fixed rater calibration. Finally, we evaluate this rule on the public ratings dataset, showing that on 857{,}256 held-out notes it recovers 75.4\% of production's CRH decisions, adds 35{,}945 publications that population-sampled raters grade on par with the notes only production publishes, and reaches shared notes a median 0.9 hours earlier. At matched coverage, its additions also grade above those of a reimplementation of X's core scorer with a relaxed threshold or shrinkage.

\section{Related Work}
\label{sec:related}

\subsection{Community Notes in practice}
A growing body of literature has looked at community fact-checking in deployment, with many focusing on X's Community Notes given their open release of code and data \citep{prollochs2022community, pilarski2024snoping, drolsbach2023diffusion}. These studies found that fact-checkers and Birdwatch notes (now Community Notes) concur far more often than they disagree \citep{saeed2022crowd}, and medical professionals judged 98\% of COVID-19 vaccine notes on X to be accurate \citep{allen2024covid}. Notes also regularly include high-quality sources for proof \citep{borenstein2025replace, solovev2025sources}. Studies also pointed to timing as a drawback, since notes historically become visible after roughly 80\% of a post's engagement has already occurred \citep{renault2024, chuai2024}, and the staggered roll-out of the program produced limited aggregate reductions in misinformation engagement \citep{chuai2024rollout}. We seek to improve the selection rule itself to address these findings.

\subsection{LLMs for corrective content}
LLMs are used to identify claims to be checked, find appropriate evidence, and produce verdicts \citep{guo2022survey}, but they also make creating persuasive false content easier \citep{pan2023pollution, chen2024combating}. Within Community Notes, AI note writers have been authorized to operate on the site since July 2025, allowing external users to run their own AI pipeline on the site \citep{li2026aifactcheckingwildfield}. \citet{de2025supernotes} also synthesizes consensus ``supernotes'' from existing candidates and scores them with a newly trained helpfulness model, motivated by their finding that for 91\% of posts with at least one proposed note, no note is ever shown. While this work focuses on generating more notes, we instead focus on making good notes more likely to appear.

\subsection{Manipulation and social choice foundations}
Contributors have been found to disproportionately flag and rate content they disagree with \citep{allen2022birds}, simulations show that a coordinated minority of 5-20\% of raters can suppress targeted helpful notes \citep{truong2025}, and strategically positioned accounts can manipulate the bridging algorithm by feigning cross-partisan agreement \citep{gaming2026consensus}. Such behaviour is hard to anticipate because bridging-based ranking has no formal characterisation of what it guarantees \citep{chooi2026commonground}, leaving its properties to be established empirically. Our rule states its guarantees as axioms, making them inspectable. In our rule design, we draw on anonymity and monotonicity and their role in May's characterization of majority rule~\citep{may1952majority}, and on the bribery~\citep{faliszewski2009bribery, faliszewski2016control} and margin-of-victory~\citep{magrino2011margin} literatures, which motivate our opposition margin.

\subsection{Alternative selection rules}
Other works have looked at alternatives for the scorer. \citet{qsmf2026} adds a per-rater quality parameter for sample efficiency and manipulation resistance. \citet{truong2025} proposes filtering unreliable raters, and X itself has both tightened rating thresholds \citep{arjmandilari2025} and, since late 2025, deployed a second learned model, the Gaussian scorer, that publishes lower-rating-volume notes after Core has a first look. All of these focus on retraining or replacing the scoring model. We instead leave the scorer intact and build on top of it, publishing notes with evidence the scorer previously deemed sufficient and reusing the rater model and decision history while removing the dependence on rating volume (\S\ref{sec:mf}). \S\ref{sec:results} additionally compares this rule with loosened reimplementations of the scorer.

\vspace{-5pt}
\section{Methods}
\label{sec:Method}
\vspace{-5pt}
We propose a guarded precedent-closure selection rule for crowd-sourced bridging algorithms. Our criterion publishes more notes and does so sooner, while certifying each admission against the platform's publication history. We first review Community Notes' current core bridging algorithm and the costs of its design choices, then formalize our rule.

\vspace{-5pt}
\subsection{Community Notes' matrix factorization}
\label{sec:mf}
\vspace{-5pt}

Community Notes' core scoring component is a matrix factorization with intercepts on $y_{un}\in\{0,0.5,1\}$ for note $n$ by rater $u$, 
\begin{equation}
\hat r_{un} = \mu + i_u + i_n + f_u f_n
\qquad
\min_{\mu,\,i,\,f}\;\sum_{(u,n)\in\mathcal R}\big(y_{un}-\hat r_{un}\big)^2
+\lambda_{\mathrm{int}}\big(\mu^2+i_u^2+i_n^2\big)
+\lambda_{f}\big(f_u^2+f_n^2\big)
\label{eq:mf}
\end{equation}
fit by regularized least squares over the full rating matrix (${\sim}3$M notes $\times$ ${\sim}1.5$M raters, ${\sim}9$M parameters) and refit hourly \citep{wojcik2022}. Crucially, the penalty term is asymmetric, as intercepts are penalized at $\lambda_{\mathrm{int}}=0.15$, five times the factor penalty $\lambda_f=0.03$, with the stated goal of capturing a cleaner signal of true helpfulness through $i_n$, thus raising the bar on the evidence required to publish.

For publication (marked by the Currently Rated Helpful (CRH) status), a note must reach $i_n \ge 0.40$ and $|f_n| < 0.5$, along with additional gates (minimums on a note's helpful-minus-not-helpful ratings, along with other constraints). Once a note satisfies all conditions to publish, it is held in a preview state for a short time before its public status visibly changes, and its status can continue to change until 2 weeks after creation, when it is permanently locked. As such, we differentiate between a note's decision time (the moment it passes all gates to become CRH), its status time (when it publicly becomes CRH and appears on users' feeds), and its lock time.

The scorer's classification as a bridging algorithm comes from its $f$ terms. Predictable agreement between a rater and a note's ``point of view'' is absorbed by $f_u$ (the latent worldview of a rater) and $f_n$ (the worldview of the note), while perceived helpfulness uncorrelated to a rater's mean generosity or societal leaning is captured by $i_n$. A core assumption of this model is therefore that raters can be partitioned into two poles ($\mathrm{sign}(f_u)$), with $|f_u|$ measuring how far along the axis they stand.

The intercept penalty is the main barrier to CRH status. Under the current normalization, the intercept ridge implies every note starts with $\tilde\lambda = \lambda_{\mathrm{int}} R/N \approx 8.97$ unbiased ghost ratings of zero helpfulness. Clearing the $i_n \ge 0.40$ threshold therefore doubles as both a quality and a volume threshold, effectively waiting for clearer signal before publishing a note. This guards against admitting notes on noise at low rating counts and against cheap coordinated support. The asymmetric penalty ($\lambda_{\mathrm{int}}=5\lambda_f$) also encodes an attribution assumption: support correlated with rater factor is preferentially absorbed by $f_uf_n$ rather than $i_n$. Consequently, an additional supporting rating from a rater whose factor aligns with the note's can lower $i_n$.

A direct consequence of these design decisions is that while many notes have a satisfactory support to opposition ratio, many simply do not have enough ratings to clear the volume bar. For a note with $k$ ratings, writing $\bar r=\tfrac1k\sum_u(y_{un}-\mu-i_u-f_uf_n)$ for its mean residual per rating, we have
\begin{equation}
i_n \;=\; \bar r\cdot\frac{k}{k+\tilde\lambda},
\qquad\text{so}\qquad
i_n\ge 0.40 \;\Longleftrightarrow\; \bar r\ \ge\ 0.40\big(1+\tilde\lambda/k\big)
\label{eq:asymptote}
\end{equation}

\begin{figure}[H]
\centering
\includegraphics[width=0.65\textwidth]{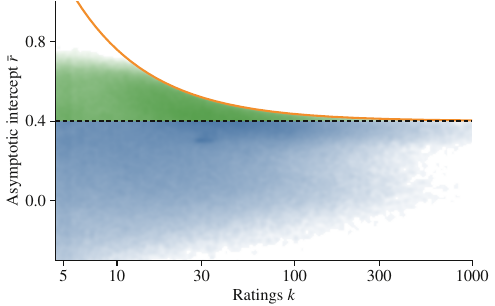}
\caption{Need More Ratings notes that clear every Core gate except $i_n\ge0.40$, by rating count $k$ and asymptotic intercept $\bar r$. The dashed line is the unshrunk bar $\bar r=0.40$, while the orange is the gate actually applied, $\bar r\ge0.40\,(1+\tilde\lambda/k)$. The notes shaded in green are those whose ratings clear the bar asymptotically but not yet the volume needed.}
\label{fig:waiting}
\end{figure}

If future ratings resemble those already received, then $\bar r$ represents the asymptote of a note's score. Figure~\ref{fig:waiting} plots $\bar r$ for all NMR (Need More Ratings) notes that pass every Core gate except $i_n \geq 0.4$ in our converged reimplementation of Core, refit on all ratings (Appendix~\ref{app:baselines}). The green region represents notes whose asymptote clears the $\ge 0.4$ boundary but sit under the volume-controlled intercept gate, therefore waiting for more ratings of the same quality to become CRH. Of the 529{,}847 notes plotted, 178{,}783 (33.7\%) are in this region. Among the 152{,}749 of these created at least 180 days before the data snapshot, whose full rating window we observe, 44.3\% received no further ratings from earned-in raters after their first day, and 68.8\% none after three days. The rule in \S\ref{sec:ourmodel} targets this volume dependence.

\subsection{Guarded precedent-closure rule}
\label{sec:ourmodel}
\vspace{-5pt}
To motivate the adoption of our rule, we rely on 2 design principles. First, instead of defining new assumptions that may collide with those of a previous bridging algorithm, we derive a publication threshold that selects a note only if its evidence coordinate-wise dominates evidence the platform's own decision history previously deemed sufficient. We call this \emph{coordinate-wise precedent certification} (\S\ref{sec:gp}): every admission dominates $K$ historical acceptances on the chosen evidence coordinates. Second, we rely on learned quantities to describe raters, but not the notes being judged (so metrics like $\mu$, $i_u$, and $\mathrm{sign}(f_u)$ can be used, but not $i_n$ and $f_n$) to retain the factorization used to qualify raters, while also avoiding the costs of shrinkage and reattribution detailed in \S\ref{sec:mf}. All rater quantities ($q_u$, $\mathrm{sign}(f_u)$, the far band and the caps $c_H$, $c_O$) come from one factorization fit to convergence on ratings before the evaluation cutoff (Appendix~\ref{app:coords}).

Our rule summarizes a note by eight coordinates: calibrated support and opposition per pole $(\mathrm{WH}_+,\mathrm{WH}_-)$, $(\mathrm{WU}_+,\mathrm{WU}_-)$, raw rating counts per pole $(n_+,n_-)$, and far-rater counts $(f_+,f_-)$.

\begin{wraptable}{r}{0.40\textwidth}
  \vspace{-\dimexpr\intextsep+5pt\relax}
  \centering
  \small
  \setlength{\tabcolsep}{6pt}
  \caption{Rater camp $\times$ direction.}
  \label{tab:coords}
  \begin{tabular}{@{}lcc@{}}
    \toprule
    raters    & support   & opposition \\
              & $y > q_u$ & $y < q_u$  \\
    \midrule
    $f_u > 0$ & $WH_+$    & $WU_+$     \\
    $f_u < 0$ & $WH_-$    & $WU_-$     \\
    \bottomrule
  \end{tabular}
  \vspace{-\dimexpr\intextsep-2pt\relax}
\end{wraptable}

The first four weights measure the evidence of ratings. We define a rating's residual as $y_{un}-q_u$, where $q_u$ is a leave-one-out empirical Bayes estimate of the rater baseline $\mu+i_u$. This captures how helpful a rater finds a note beyond their baseline generosity, with residuals being capped at the 90th percentile of strictly positive pre-cutoff residuals ($c_{\mathrm H}$ for support, $c_{\mathrm O}$ for opposition) so no single account carries outsized weight. We then separate raters by pole and sum their weights to get our first 4 metrics, as described in Table~\ref{tab:coords}. These are capped, baseline-adjusted analogues of the factorization's intercept evidence (\S\ref{sec:mf}), but un-shrunk, without the $f_uf_n$ reattribution, and signed by pole.

The remaining four are counts to improve robustness. Since two raw ratings that fall far from their rater's baseline generosity could outweigh five moderate ones, we introduce raw per-pole counts $n_\pm$, which count how many ratings a note received from raters with $f_u > 0$ and $f_u < 0$ separately. Finally, the far counts $f_\pm$ count the number of ratings from raters with $|f_u|\ge 0.35$ for each pole, since raters who most strongly disapprove of a note frequently sit at the far ends of the factor spectrum, and waiting for their signal is very informative. Both of these guards are ablated in \S\ref{sec:ablations}.

Our decision rule then simply becomes a comparison against history. We say evidence vector $x_a$ dominates $x_b$ if $x_a$ is at least as strong on all eight coordinates: at least as much support weight, headcount, and far exposure on each pole, and at most as much opposition weight. The precedents we compare new notes against are the notes that have already been granted CRH status by the system, frozen at their decision-time coordinates. Similarly, we also freeze CRNH notes (Currently Rated Not Helpful status) as our anti-precedents. Precedents are evidential rather than topical: dominating a precedent certifies that the platform has already published a note on weaker rater evidence on every coordinate, whatever that note said. The content is still judged by the raters, while the coordinates summarize how strongly and how broadly they endorsed it. A candidate is therefore published iff: (1) it dominates at least $K$ precedents, (2) no single anti-precedent dominates it, (3) $\mathrm{WH}_+ + \mathrm{WH}_-$ clears a floor of $3.0$ ($\approx 8$ typical ratings), and (4) conditions (1)--(3) still hold after one maximum-weight opposition rating ($c_{\mathrm O}$) is added to either pole.

\begin{figure}[t]
\centering
\includegraphics[width=\linewidth]{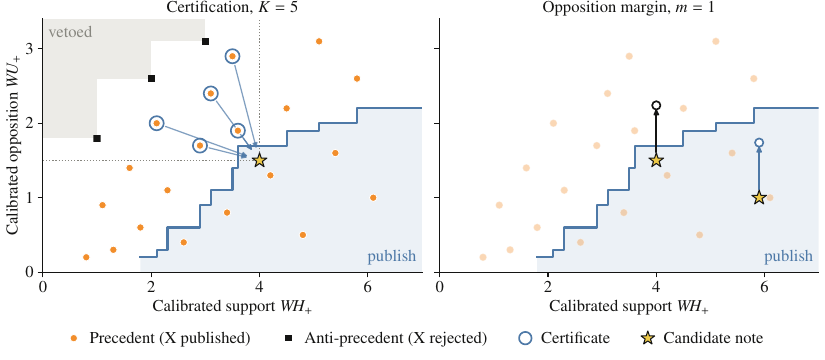}
\vspace{-10pt}
\caption{\textbf{Left:} An admission and rejection boundary at $K=5$ given simulated precedents. \textbf{Right:} Adding an extra maximum-weighted opposition rating to test robustness.}
\label{fig:rule}
\end{figure} 

Figure~\ref{fig:rule} demonstrates this geometry at $K=5$. On the left, the candidate (star) dominates the five circled precedents while staying outside of the grey vetoed region, offering more support and less opposition than 5 previously accepted notes. On the right, the margin test adds one maximum-weight opposition rating (in this case, $+c_{\mathrm O}$). A candidate with slack $0.20$ is pushed out of the region and must wait for more ratings, while one with slack $1.20$ survives and receives CRH status. Every publication is thereby certified by $K$ named precedents, plus the guarantee that, with calibration and ledgers fixed, one additional capped opposition rating cannot invalidate its base-region certificate.

\subsection{Properties and guarantees}
\label{sec:gp}

We establish the theoretical guarantees for the guarded precedent rule with fixed precedents, rater calibrations, and eligibility, dwell, and lock gates. For compactness, write $H_s,U_s,N_s,F_s$ for $\mathrm{WH}_s,\mathrm{WU}_s,n_s,f_s$ for each $s\in\{+,-\}$. For each note $n$, express its evidence vector as:
\[
x_n=(H_+,H_-,U_+,U_-,N_+,N_-,F_+,F_-)
\]
We write $x\succeq z$ if evidence vector $x$ dominates $z$ (from \S\ref{sec:ourmodel}). This is equivalent to:
\begin{equation}
x\succeq z
\quad\Longleftrightarrow\quad
\begin{aligned}
H_s(x)&\ge H_s(z), & U_s(x)&\le U_s(z),\\
N_s(x)&\ge N_s(z), & F_s(x)&\ge F_s(z),
\end{aligned}
\qquad s\in\{+,-\}
\label{eq:evidence-order}
\end{equation}
Let $\mathcal P$ and $\mathcal A$ be the set of precedent and anti-precedent notes. For any candidate vector $x$, define:
\[
d_{\mathcal P}(x)
=
\bigl|\{p\in\mathcal P:x\succeq x_p\}\bigr|
\]
the number of precedents whose frozen evidence vectors are dominated by $x$. Before the opposition-margin guard, the eligible region for publication is therefore:
\begin{equation}
\mathcal E_K(\mathcal P,\mathcal A)
=\left\{x:
d_{\mathcal P}(x)\ge K,\;
\nexists a\in\mathcal A\text{ with }x_a\succeq x,\;
H_+(x)+H_-(x)\ge h_0
\right\}
\label{eq:base-region}
\end{equation}
where $K\in\mathbb N$ and $h_0=3.0$ in our implementation. For $m\in\mathbb N_0$ and $c_{\mathrm O}\ge0$, set
\begin{equation}
\Lambda_m
=\{(\ell_+,\ell_-)\in\mathbb N_0^2:\ell_++\ell_-\le m\},
\qquad
T_\ell x
=x+(0,0,\ell_+c_{\mathrm O},\ell_-c_{\mathrm O},0,0,0,0)
\label{eq:stress-map}
\end{equation}
So $T_\ell$ adds at most $m$ maximum-weight opposition ratings across the two poles, while granting no favorable headcount or far-rater credit. The opposition guarded region is therefore:
\begin{equation}
\mathcal G_{K,m}(\mathcal P,\mathcal A)
=\bigcap_{\ell\in\Lambda_m}
\left\{x:T_\ell x\in
\mathcal E_K(\mathcal P,\mathcal A)\right\}
\label{eq:guarded-region}
\end{equation}
Our default rule uses $m=1$.

\begin{lemma}[Order structure]
\label{lem:order-structure}
For fixed $\mathcal P$, $\mathcal A$, and calibration, $\mathcal G_{K,m}(\mathcal P,\mathcal A)$ is an up-set under $\succeq$: if $x$ is admitted and $z\succeq x$, then $z$ is admitted.
\end{lemma}
\vspace{-0.8em}
\begin{proof}[Proof sketch]
Dominance is transitive, so $z$ inherits every precedent dominated by $x$. If an anti-precedent dominated $z$, it would also dominate $x$. The support floor is increasing under $\succeq$, and every stress map is order-preserving, so $z\succeq x$ implies $T_\ell z\succeq T_\ell x$. Each condition in \eqref{eq:guarded-region} is therefore an up-set, as is their intersection. Appendix~\ref{app:gp-proofs} gives the complete proof.
\end{proof}

\begin{corollary}[Coherence, monotonicity, and history updates]
\label{cor:order-consequences}
Under the conditions of Lemma~\ref{lem:order-structure}:
\begin{enumerate}
    \item[(a)] \emph{Dominance coherence:} a candidate that dominates an admitted candidate cannot be rejected by the coordinate rule,
    \item[(b)] \emph{Support-addition monotonicity:} adding a rating with $y_{un}>q_u$ cannot revoke admission,
    \item[(c)] \emph{Two-sided history monotonicity:} adding a qualifying precedent weakly expands the admission region, whereas adding an anti-precedent weakly contracts it,
    \item[(d)] \emph{Guard nesting:} increasing $m$ weakly contracts the admission region.
\end{enumerate}
\end{corollary}

For (b), a supporting rating increases one $H_s$ and $N_s$, may increase $F_s$, and leaves opposition unchanged. Part (c) follows because enlarging $\mathcal P$ can only increase $d_{\mathcal P}$, whereas enlarging $\mathcal A$ can only add vetoes. Part (d) follows from $\Lambda_m\subseteq\Lambda_{m+1}$. Full set-inclusion arguments appear in Appendix~\ref{app:gp-proofs}.

\begin{lemma}[Certified opposition margin]
\label{prop:opposition-margin}
Suppose each eligible added rating contributes at most $c_{\mathrm O}$ opposition weight to one pole, and otherwise only nonnegative increments to support, headcount, and far exposure. If $x\in\mathcal G_{K,m}$, every addition of at most $m$ such ratings leaves the note in $\mathcal E_K$. Hence more than $m$ added ratings are required to invalidate its base-region decision:
\begin{equation}
x\in\mathcal G_{K,m+\ell}
\quad\Longrightarrow\quad
x^{\prime}\in\mathcal G_{K,m},
\label{eq:margin-composition}
\end{equation}
where $x'$ is the coordinate vector after at most $\ell$ such additions.
\end{lemma}

Since the stress test assigns each hypothetical rating
the maximum opposition weight without counting headcount and far-rater exposure, it is conservative. Thus, if $\rho$ records the number of added ratings on each pole, the actual post-attack vector $x'$ is at least as strong as the simulated vector $T_\rho x$. An admitted note therefore remains in the base region $\mathcal E_K$ after at most $m$ added ratings and still dominates $K$ named precedents under each tested placement. Appendix~\ref{app:gp-proofs} gives the full proof and attack-model qualifications.

\paragraph{Contrast with the normalized gate.} The incumbent's gate lacks monotonicity. Writing its slack as $\Delta=\sum_j e_j-t(k+\tilde\lambda)$ with $t=0.40$, where $e_j$ are the note's residuals, adding a supporting rating with residual $e$ changes the slack to $\Delta+e-t$, because the rating increments the shrunk count as well as the residual sum. Any supporting rating with $e<t$ therefore counts against the note, and reverses a near-boundary acceptance exactly when $0\le\Delta<t-e$. Corollary~\ref{cor:order-consequences}(b) rules this out by thresholding unnormalized sums.

\paragraph{Deployment.} Precedent backing is not intrinsic set containment: with $K>1$ or either guard, a historical precedent need not itself re-qualify (Appendix~\ref{app:no-containment}). The rule therefore keeps previously accepted notes published and does not re-evaluate them; \S\ref{sec:discussion} discusses re-checking additions until their status locks, as X does for its own statuses. Deployed as the promotion-layer union $D_{\mathrm{incumbent}}\lor D_{\mathrm{ours}}$, incumbent publications are preserved by construction, while each additional publication carries the named precedent witnesses above.

\section{Results}
\label{sec:results}
\begin{figure}[t]
\centering
\includegraphics[width=\textwidth]{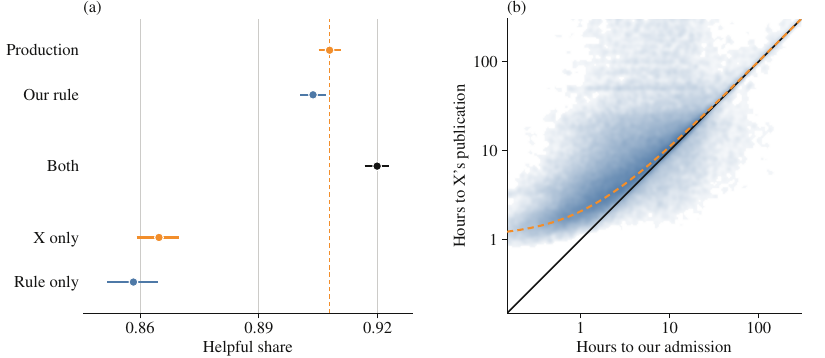}
\caption{(a) Helpful share among population-sampled raters. \emph{X only} and \emph{Rule only} are the notes on which the two systems disagree, while \emph{Both} publish is their overlap. The dashed line marks production. (b) Hours to our admission vs to X's publication on notes both publish, on log scales. Notes above the solid parity line are admitted by our rule before X publishes them, and notes above the dashed orange line stay ahead even after we add X's 65-minute ingest delay and hold to our rule.}
\label{fig:quality}
\end{figure}

\begin{wraptable}{r}{0.45\textwidth}
\vspace{-\intextsep}
\vspace{-7pt}
\centering
\small
\setlength{\tabcolsep}{3.5pt}
\renewcommand{\arraystretch}{0.98}
\caption{Precedent rule ($K{=}5$, $m{=}1$) vs.\ X's production scorers on 857,256 held-out notes.}
\label{tab:coverage}
\vspace{4pt}
\begin{tabular}{@{}lrr@{}}
\toprule
& \multicolumn{1}{c}{X} & \multicolumn{1}{c}{Ours} \\
\midrule
\multicolumn{3}{@{}l}{\textit{Publication}} \\
Notes published & 91,670 & 105,019 \\
\quad this rule only & 22,596 & 35,945 \\
\quad not firm-rejected & -- & 23,047 \\
\midrule
\multicolumn{3}{@{}l}{\textit{This rule only, medians}} \\
Ratings & 43 & 50 \\
Minority-pole ratings & 12 & 12 \\
$<3$ minority-pole ratings & 3.6\% & 0.3\% \\
\midrule
\multicolumn{3}{@{}l}{\textit{Both publish (69,074 notes)}} \\
Median lead (hours) & -- & 0.9 \\
Mean lead (hours) & -- & 9.3 \\
Published first & 31.3\% & 68.7\% \\
\bottomrule
\end{tabular}
\vspace{-\intextsep}
\vspace{2pt}
\end{wraptable}

We test the implementation of the guarded precedent rule using X's publicly available datasets of raters and notes. We take all notes written before 2025-11-10 as our precedents and anti-precedents (the date the ``Gaussian scorer'', a new addition to X's algorithm, entered production), and test on the held out set of 857,256 notes created between 2025-11-10 and 2026-08-30. We compare against the decisions X's production algorithm recorded over the same period. We exclude population-sampled ratings so they can later be used to evaluate the usefulness of notes, and our rater calibration is fit to convergence on ratings from before the cutoff only.

We find that the new rule publishes 105,019 of these notes against production's 91,670, recovering 75.4\% of what X published and adding 35,945 notes it did not, 23,047 of which Core does not flag as firm-rejected. Combined with production's CRHed notes, the rule therefore publishes 39.2\% more notes (25.1\% counting only additions Core does not firm-reject), with only 31 of its additions being notes X eventually rated \textsc{not helpful}. These added notes carry slightly fewer ratings than X's publications overall (median 50 against 57), but have comparable opposition shares: the median share of ratings marked \textsc{not helpful} is 6.4\% for our additions against 6.7\% for X's.

X also publishes 22,596 notes that our rule does not admit. These are more contested, with a median of 15.0\% of ratings marked \textsc{not helpful}, and remain published when our rule is deployed as a union with production. Beyond note additions, our rule also publishes notes faster. Since X's CRH timestamps mark when a note actually went live, we charge our rule for X's delay in getting ratings into the scorer and for the temporary hold applied before a status is shown, estimated at 20 and 45 minutes (the mean of X's stated 30- and 60-minute bounds). Even then, of the 69,074 notes both systems publish, our rule publishes first on 68.7\%, a median 0.9 hours earlier (Figure~\ref{fig:quality}b). The mean lead, 9.3 hours, is far larger than the median because of production's tail: its 90th percentile is 35.6 hours against our 18.5.

\begin{figure}[t]
\centering
\includegraphics[width=\textwidth]{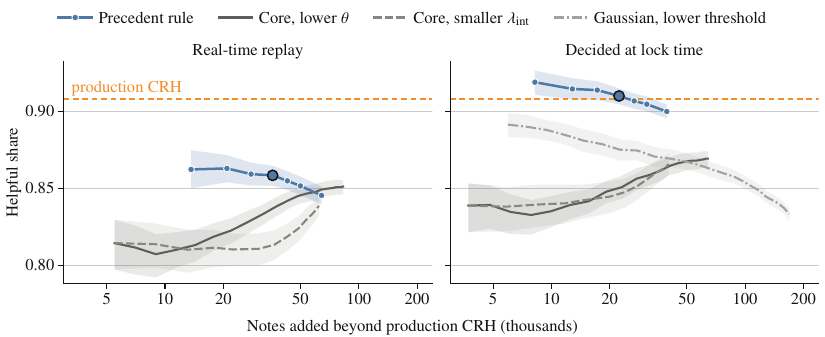}
\caption{Coverage against quality of the notes each rule adds beyond production, in replay (left) and decided at lock time (right). Blue: our rule, $K$ from 50 to 1 (ringed: $K=5$); grey: Core with a lower intercept threshold (0.40 to 0.20) or a smaller intercept penalty (0.15 to 0.005). Gaussian (right): Gaussian-scorer, threshold lowered from 0.62 to 0.40 (production: 0.51). Bands: 95\% intervals.}
\label{fig:frontier}
\end{figure}

We now look into the quality of the new notes our rule accepts. Since users who rate a note must have been shown the original post by their feed, they are likely to have some preexisting bias about the topic. To get more objective rankings, Community Notes occasionally shows users notes at random and marks the resulting ratings ``population\_sampled'' in the public data. For each set of notes, we report the share of helpful ratings among its helpful and not-helpful population-sampled ratings, pooled over ratings rather than averaged over notes, for notes with at least three such ratings. On the notes where the two systems disagree, our additions grade 0.858 against 0.865 for the notes only X publishes, a difference within noise (Figure~\ref{fig:quality}a), while notes both systems publish grade 0.920. If decisions instead wait until a note's status locks, the rule still adds 22,351 notes, which grade 0.910, on par with production's own publications at 0.908.

\begin{figure}[t]
\centering
\includegraphics[width=0.85\textwidth]{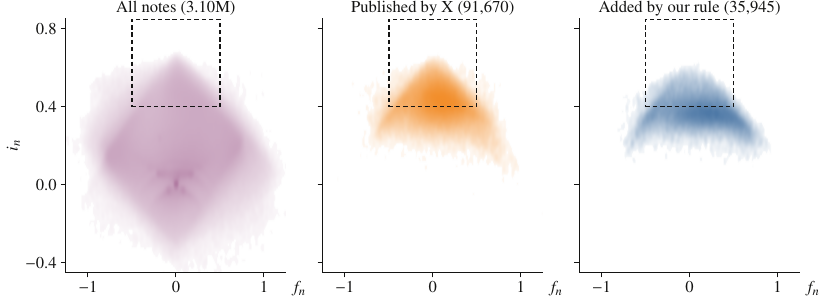}
\caption{Notes on $f_n$ against $i_n$ in our converged Core reimplementation: all notes (left), notes X publishes (middle) and notes our rule adds (right). Dashed box: Core's publication gate ($i_n\ge0.40$, $|f_n|<0.5$). Shading: note density.}
\label{fig:mfspace}
\end{figure}

To see where these additions sit in Core's own model, Figure~\ref{fig:mfspace} places every note at its factor $f_n$ and intercept $i_n$ in our converged reimplementation of Core, with Core's gate drawn as a box. X's publications sit mostly inside it, with 59.2\% clearing both conditions (median $i_n$ of 0.416), while our additions form the same shape displaced downwards, with 73.6\% of them outside the gate. Our rule never reads $f_n$, yet its additions stay centered near $f_n=0$ like X's publications, suggesting that requiring evidence from both poles recovers much of what Core's $|f_n|<0.5$ condition enforces.

It is also worth considering whether relaxing production's own scoring would publish the same notes. Figure~\ref{fig:frontier} compares our rule at matched coverage with reimplementations of Core whose intercept threshold is lowered or whose intercept penalty is shrunk. These relaxations pick largely different notes from ours (Jaccard 0.30--0.40), and the notes only our rule adds grade higher than the ones only they add: by 0.034 and 0.106 in replay, and by 0.079 and 0.097 when decided at lock time (more in Appendix~\ref{app:baselines}). In terms of hyperparameters, choosing $K$ mainly sets coverage and speed rather than quality, as from $K=2$ to $50$, the helpful share stays between 0.851 and 0.863 while the number of additions falls from 50,002 to 13,632 (Appendix~\ref{app:sens}). The opposition margin, however, matters more. Without it ($m=0$), the rule adds 139,019 notes grading 0.827, while $m=2$ gives up the speed advantage without improving quality, which is why we choose $m=1$.

\label{sec:ablations}
Finally, we run some additional tests to validate the necessity of our additional metrics. At admission, our count and far guards leave only 1.2\% of additions with fewer than three ratings on a pole and 6.9\% without a far-band rating on one, against 2.2\% and 7.3\% of X's own publications, and 46.4\% and 42.3\% without these guards. The guard that filters out most notes, however, remains our robustness test, which removes 53.3\% of the notes that reach it.

\section{Discussion}
\label{sec:discussion}
\vspace{-5pt}
X's current publication gate is both a quality bar and a volume bar: because the intercept it thresholds is shrunk, clearing $i_n \ge 0.40$ demands total evidence that grows with every rating a note receives. As such, replacing this threshold with an order over the platform's own published history recovers notes that simply had not yet crossed the volume bar. Deployed as a union with production, the rule adds 35,945 notes that grade on par with the notes only production publishes, and reaches shared notes a median 0.9 hours earlier. Re-checking these additions until their status locks, as X already does for its own statuses, would keep 22,351 of them, grading 0.910, on par with production's own publications. In other words, the rule publishes notes held back by a bar that also scales with attention, and certifies each addition against $K$ historical acceptances.

\section{Limitations and Future Work}
\label{sec:limitations}
\vspace{-5pt}
It is important to note that X's current publication stack is not a single algorithm, and that we read the decisions of its added modules, such as the Gaussian scorer, from the published data rather than rerunning them independently. And while all modules were introduced before the start of our testing set, the newest modules had little time to act on newer notes, so X's recorded decisions may understate what its current stack would publish.

Our additions still grade below production's publications overall on population-sampled ratings (0.858 against 0.908), and 34.1\% of those admitted before lock fail the rule on lock-time evidence (grading 0.826) but stay published unless re-checked. Layered over production, the rule also makes coordinated suppression harder but promotion easier: one established account suppresses 5.6\% of targets instead of 24.4\%, while five promote 17.7\% instead of 4.0\%. A support-removal margin narrows this gap at some cost in coverage and speed (Appendix~\ref{app:attack}). Our comparison methods are also reimplementations that reproduce Core only moderately (Appendix~\ref{app:baselines}), and against a loosened reimplementation of X's Gaussian scorer, the notes it alone adds grade 0.021 [0.002, 0.040] higher than ours. Our calibration is also fit once before the cutoff, so a deployment would need to refit it, and offline we cannot measure how publishing changes the ratings a note receives.

For future work, we hope to collaborate with X's Community Notes team on adapting this rule for the production stack, and to apply it to other bridging-based note systems, which only need ratings, a rater axis and a record of past decisions. Precedents could also be weighted by semantic relevance, for instance from text embeddings; since such weights depend only on the texts, Lemma~\ref{lem:order-structure} and Lemma~\ref{prop:opposition-margin} would still hold for each candidate.

\bibliographystyle{plainnat}
\bibliography{example}

\newpage
\appendix
\section{Proofs for the properties and guarantees}
\label{app:gp-proofs}

\subsection{Coordinate map and standing assumptions}
\label{app:coords}

Let $\mathcal U^*$ denote the set of \emph{coordinate-eligible} raters, which we define as raters with a fitted baseline $q_u$ and a fitted factor $f_u\neq 0$ at the calibration cutoff. Note that raters who first rate after the cutoff have no fitted factor and are outside $\mathcal U^*$. Fix $\tau=0.35$ and assign each $u\in\mathcal U^*$ the calibration type
\[
\theta_u=\bigl(q_u,\ \operatorname{sign}(f_u),\ \mathbf 1\{|f_u|\ge\tau\}\bigr).
\]
For an observed rating $y_{un}$ define the capped residual weights
\begin{equation}
h_{un}=\min\{(y_{un}-q_u)_+,\,c_{\mathrm H}\},
\qquad
o_{un}=\min\{(q_u-y_{un})_+,\,c_{\mathrm O}\},
\label{eq:capped-residuals}
\end{equation}
so that at most one of $h_{un},o_{un}$ is nonzero. Writing $s=\operatorname{sign}(f_u)$ and $\varphi_u=\mathbf 1\{|f_u|\ge\tau\}$, the per-rating coordinate map is:
\begin{equation}
\phi(y_{un},\theta_u)=
\bigl(h_{un}\mathbf 1_{s=+},\ h_{un}\mathbf 1_{s=-},\
o_{un}\mathbf 1_{s=+},\ o_{un}\mathbf 1_{s=-},\
\mathbf 1_{s=+},\ \mathbf 1_{s=-},\
\varphi_u\mathbf 1_{s=+},\ \varphi_u\mathbf 1_{s=-}\bigr),
\label{eq:rating-coordinate-map}
\end{equation}
with the convention $\phi(\varnothing,\theta_u)=0$ when $u$ has not rated $n$. The evidence vector of note $n$ is
\[
x_n=(H_+,H_-,U_+,U_-,N_+,N_-,F_+,F_-)=\sum_{u\in\mathcal U^*}\phi(y_{un},\theta_u),
\]
which covers the eight coordinates of \S\ref{sec:ourmodel} (there written $\mathrm{WH}_\pm,\mathrm{WU}_\pm,n_\pm,f_\pm$). A rating with $y_{un}=q_u$ contributes no weight but counts toward $N_s$ and, if $|f_u|\ge\tau$, toward $F_s$, and ratings from raters outside $\mathcal U^*$ contribute nothing. Every entry of $\phi$ is nonnegative.

All results in this appendix hold under the following standing assumptions, which together constitute the threat model.
\begin{itemize}
\item[(A1)] \emph{Fixed calibration.} $q_u$, $\operatorname{sign}(f_u)$, $\mathbf 1\{|f_u|\ge\tau\}$, $c_{\mathrm H}$, and $c_{\mathrm O}$ are fitted once, on ratings before the cutoff, and are not refit within the horizon of the claim.
\item[(A2)] \emph{Fixed ledgers.} $\mathcal P$ and $\mathcal A$ are fixed sets of precedent and anti-precedent notes whose vectors are frozen at their decision time.
\item[(A3)] \emph{Append-only ratings.} Ratings on the candidate are only added, never revised or deleted, and each added rating comes from a rater who has not previously rated the note.
\item[(A4)] \emph{Orthogonal gates.} Eligibility, dwell, and lock gates are applied independently of the coordinates and are not manipulated.
\end{itemize}
Throughout, $\succeq$ is the product order \eqref{eq:evidence-order}; it is reflexive, transitive, and antisymmetric. A set $S$ is an \emph{up-set} if $x\in S$ and $z\succeq x$ imply
$z\in S$.

\subsection{Proof of Lemma~\ref{lem:order-structure}}

\begin{proof}
We show that each of the sets whose intersection defines $\mathcal G_{K,m}(\mathcal P,\mathcal A)$ in \eqref{eq:guarded-region} is an up-set.

\emph{Step 1: $\mathcal E_K(\mathcal P,\mathcal A)$ is an up-set.} Let $x\in\mathcal E_K$ and $z\succeq x$. (i) If $x\succeq x_p$ for a precedent $p$, transitivity gives $z\succeq x_p$, so $\{p:x\succeq x_p\}\subseteq\{p:z\succeq x_p\}$ and $d_{\mathcal P}(z)\ge d_{\mathcal P}(x)\ge K$. (ii) If some $a\in\mathcal A$ satisfied $x_a\succeq z$, then $x_a\succeq z\succeq x$ would give $x_a\succeq x$, contradicting $x\in\mathcal E_K$; hence no anti-precedent dominates $z$. (iii) $z\succeq x$ implies $H_+(z)\ge H_+(x)$ and $H_-(z)\ge H_-(x)$, so $H_+(z)+H_-(z)\ge H_+(x)+H_-(x)\ge h_0$. Thus $z\in\mathcal E_K$.

\emph{Step 2: each stress map preserves $\succeq$.} For fixed $\ell$, $T_\ell$ is a translation by the constant vector $(0,0,\ell_+c_{\mathrm O},\ell_-c_{\mathrm O},0,0,0,0)$, so $T_\ell z-T_\ell x=z-x$ coordinatewise. Since \eqref{eq:evidence-order} depends only on the signs of coordinate differences, $z\succeq x$ if and only if $T_\ell z\succeq T_\ell x$.

\emph{Step 3: conclusion.} Let $x\in\mathcal G_{K,m}$ and $z\succeq x$. For every $\ell\in\Lambda_m$ we have $T_\ell x\in\mathcal E_K$ by definition, $T_\ell z\succeq T_\ell x$ by Step~2, and therefore $T_\ell z\in\mathcal E_K$ by Step~1. Hence $z\in\mathcal G_{K,m}$. \end{proof}

\subsection{Proof of Corollary~\ref{cor:order-consequences}}

\begin{proof}
\emph{(a) Dominance coherence.} If $x\in\mathcal G_{K,m}$ is admitted and a candidate has vector $z\succeq x$, then $z\in\mathcal G_{K,m}$ by Lemma~\ref{lem:order-structure}; the coordinate rule therefore cannot reject it. (Rejection by a gate outside the coordinate rule is excluded by (A4).)

\emph{(b) Support-addition monotonicity.} Let $x$ be admitted and let a rating $y_{un}>q_u$ be appended by a rater $u$ who has not rated $n$ (A3). If $u\notin\mathcal U^*$ the vector is unchanged and admission is unaffected. Otherwise, with $s=\operatorname{sign}(f_u)$, \eqref{eq:capped-residuals} gives $h_{un}=\min\{y_{un}-q_u,c_{\mathrm H}\}>0$ and $o_{un}=0$, so by \eqref{eq:rating-coordinate-map} the new vector is $x'=x+\phi(y_{un},\theta_u)$, where $\phi(y_{un},\theta_u)$ has $U_+=U_-=0$ and all other entries nonnegative. Hence $H_s(x')\ge H_s(x)$, $N_s(x')=N_s(x)+1$, $F_s(x')\ge F_s(x)$, and $U_\pm(x')=U_\pm(x)$, i.e.\ $x'\succeq x$. Lemma~\ref{lem:order-structure} gives $x'\in\mathcal G_{K,m}$. Assumption (A1) ensures the type $\theta_u$ used to evaluate the new rating is the one used to form $x$.

\emph{(c) Two-sided history monotonicity.} Let $\mathcal P\subseteq\mathcal P'$ with $\mathcal A$ fixed. For every $x$, $\{p\in\mathcal P:x\succeq x_p\}\subseteq \{p\in\mathcal P':x\succeq x_p\}$, so $d_{\mathcal P'}(x)\ge d_{\mathcal P}(x)$, while the veto and floor conditions are unchanged; thus $\mathcal E_K(\mathcal P,\mathcal A)\subseteq \mathcal E_K(\mathcal P',\mathcal A)$. Intersecting over the same index set $\Lambda_m$ preserves inclusion, so $\mathcal G_{K,m}(\mathcal P,\mathcal A)\subseteq \mathcal G_{K,m}(\mathcal P',\mathcal A)$. Symmetrically, if $\mathcal A\subseteq\mathcal A'$ with $\mathcal P$ fixed, every veto in $\mathcal A$ is also a veto in $\mathcal A'$, so $\mathcal E_K(\mathcal P,\mathcal A')\subseteq\mathcal E_K(\mathcal P,\mathcal A)$ and hence $\mathcal G_{K,m}(\mathcal P,\mathcal A')\subseteq\mathcal G_{K,m}(\mathcal P,\mathcal A)$.

\emph{(d) Guard nesting.} $\Lambda_m\subseteq\Lambda_{m+1}$, so the intersection in \eqref{eq:guarded-region} over $\Lambda_{m+1}$ imposes a superset of the conditions imposed over $\Lambda_m$: $\mathcal G_{K,m+1}\subseteq\mathcal G_{K,m}$. \end{proof}

\subsection{Proof of Lemma~\ref{prop:opposition-margin}}

\begin{proof}
\emph{Attack model.} Under (A1)--(A4), an attacker may append ratings to the candidate, each from a distinct rater who has not yet rated it. A rating from $u\notin\mathcal U^*$ changes nothing, so we may assume every added rating is eligible. For an eligible rating on pole $s=\operatorname{sign}(f_u)$, \eqref{eq:rating-coordinate-map} shows that it adds at most $o_{un}\le c_{\mathrm O}$ to $U_s$, nothing to $U_{-s}$, and nonnegative amounts $h_{un}\ge0$, $1$, and $\varphi_u\in\{0,1\}$ to $H_s$, $N_s$, and $F_s$ respectively. This is  the hypothesis of the lemma. Note that a rating that carries support ($h_{un}>0$, $o_{un}=0$) is a special case.

\emph{Domination of the stress profile.} Suppose $j$ eligible ratings are added, with $\rho=(\rho_+,\rho_-)$ counting how many fall on each pole ($\rho_++\rho_-=j$), and let $x'$ be the resulting vector. Comparing $x'$ with $T_\rho x$ coordinatewise, for $s\in\{+,-\}$,
\[
H_s(x')\ge H_s(x)=H_s(T_\rho x),\quad
U_s(x')\le U_s(x)+\rho_s c_{\mathrm O}=U_s(T_\rho x),
\] \[
N_s(x')\ge N_s(T_\rho x),\quad
F_s(x')\ge F_s(T_\rho x),
\]
hence $x'\succeq T_\rho x$. The stress profile is thus pessimistic: it charges every added rating the maximal opposition weight and grants no headcount or far-rater credit.

\emph{Base-region certificate.} If $x\in\mathcal G_{K,m}$ and $j\le m$, then $\rho\in\Lambda_m$ and $T_\rho x\in\mathcal E_K$ by \eqref{eq:guarded-region}. Since $\mathcal E_K$ is an up-set (Step~1 of the proof of Lemma~\ref{lem:order-structure}) and $x'\succeq T_\rho x$, we get $x'\in\mathcal E_K$. In particular $x'$ dominates the $K$ precedents witnessed by $T_\rho x$, which are named at certification time. Invalidating the base-region decision therefore requires more than $m$ added ratings.

\emph{Composition \eqref{eq:margin-composition}.} Suppose $x\in\mathcal G_{K,m+\ell}$ and at most $\ell$ ratings are added, with pole counts $\rho$, $\rho_++\rho_-\le\ell$. We must show $T_\lambda x'\in\mathcal E_K$ for every $\lambda\in\Lambda_m$. Translations compose, $T_\lambda T_\rho=T_{\lambda+\rho}$, and preserve $\succeq$ (Step~2 above), so $x'\succeq T_\rho x$ implies $T_\lambda x'\succeq T_{\lambda+\rho}x$. Since $(\lambda_++\rho_+)+(\lambda_-+\rho_-)\le m+\ell$, we have $\lambda+\rho\in\Lambda_{m+\ell}$ and therefore $T_{\lambda+\rho}x\in\mathcal E_K$. The up-set property gives $T_\lambda x'\in\mathcal E_K$. As $\lambda\in\Lambda_m$ was arbitrary, $x'\in\mathcal G_{K,m}$. \end{proof}

\paragraph{Qualifications.}
(i) The lemma certifies the \emph{base} decision $\mathcal E_K$. Membership in $\mathcal G_{K,1}$ does not by itself guarantee that the attacked profile remains in $\mathcal G_{K,1}$; by \eqref{eq:margin-composition} that stronger conclusion requires an initial $\mathcal G_{K,2}$ certificate. (ii) The following lie outside the model and are not covered: revising or deleting existing ratings, which can lower $H_s$, $N_s$, or $F_s$; shifting one's own $q_u$, pole, or far status through ratings on other notes before a refit (excluded by (A1)); changes to either ledger (A2); coalitions of more than $m$ raters; and manipulation of gates outside the coordinate rule (A4). (iii) The certificate is evaluated at decision time; whether a published note is re-evaluated as further ratings arrive is a deployment choice orthogonal to the lemma.

\subsection{Precedent certification is not set containment}
\label{app:no-containment}

Membership in $\mathcal G_{K,m}$ states that the candidate's vector dominates $K$ frozen historical CRH vectors on the eight coordinates, survives the anti-precedent veto and the support floor, and does so under each stress placement. It does not state that the previous incumbent rule, with its full set of gates, would have admitted the candidate, nor that every historical precedent would itself be admitted: with $K>1$, a precedent $p$ need not dominate $K$ \emph{other} precedents, and its frozen vector need not survive the stress map or the veto. Certification is therefore relative to the chosen coordinates and the frozen ledgers. This is why \S\ref{sec:results} reports the rule's standalone decisions and the union $D_{\mathrm{incumbent}}\lor D_{\mathrm{ours}}$ separately, and why the union preserves incumbent publications by construction rather than by re-qualification.

\subsection{Support-removal margin}
\label{app:removal}

For $r\in\mathbb N_0$ and $\ell\in\Lambda_r$, with $\Lambda_r$ as in \eqref{eq:stress-map}, let
\[
S_\ell x=x-(\ell_+c_{\mathrm H},\ \ell_-c_{\mathrm H},\ 0,\ 0,\ \ell_+,\ \ell_-,\ \ell_+,\ \ell_-),
\]
and let the rule with support-removal margin $r$ admit $x$ if and only if $S_\ell x\in\mathcal G_{K,m}(\mathcal P,\mathcal A)$ for every $\ell\in\Lambda_r$. Taking $\ell=0$ shows that this region lies inside $\mathcal G_{K,m}$; since each $S_\ell$ is a translation, Step~2 of the proof of Lemma~\ref{lem:order-structure} shows that it is an up-set, so Corollary~\ref{cor:order-consequences} carries over, and the region weakly contracts as $r$ grows.

\begin{lemma}[Promotion margin]
\label{lem:promotion-margin}
Under (A1)--(A4), suppose a note's vector $x$ is not in $\mathcal G_{K,m}$ and the note receives at most $r$ added ratings, none of them opposing ($y_{un}\ge q_u$). Then the rule with support-removal margin $r$ does not admit the resulting vector.
\end{lemma}

\begin{proof}
By \eqref{eq:rating-coordinate-map}, a non-opposing rating on pole $s$ adds at most $c_{\mathrm H}$ to $H_s$, exactly one to $N_s$, at most one to $F_s$, and nothing to $U_+$ or $U_-$. Let $\rho_s$ of the $j\le r$ added ratings fall on pole $s$, so that $\rho\in\Lambda_r$, and let $x'$ be the resulting vector. Then $H_s(x)\ge H_s(x')-\rho_s c_{\mathrm H}$, $U_s(x)=U_s(x')$, $N_s(x)=N_s(x')-\rho_s$ and $F_s(x)\ge F_s(x')-\rho_s$, so $x\succeq S_\rho x'$. If the rule admitted $x'$, then $S_\rho x'\in\mathcal G_{K,m}$, and because $\mathcal G_{K,m}$ is an up-set (Lemma~\ref{lem:order-structure}), $x\in\mathcal G_{K,m}$, a contradiction.
\end{proof}

\section{Parameter sensitivity}
\label{app:sens}

\begin{figure}[t]
\centering
\includegraphics[width=\textwidth]{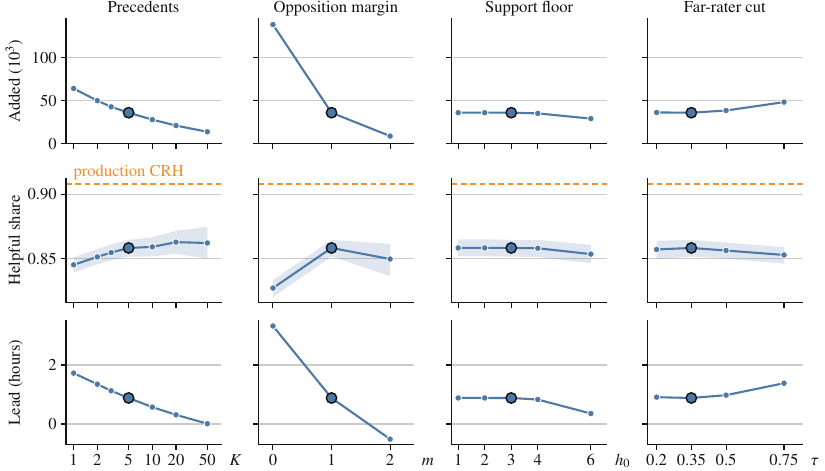}
\caption{Ablations of each parameter, with the default ringed. Columns: precedents to dominate ($K$), opposition margin ($m$), support floor ($h_0$) and far-rater cut ($\tau$). Rows: notes added beyond production; their helpful share (95\% bootstrap bands; dashed: production); median lead over production on notes both publish.}
\label{fig:sensitivity}
\end{figure}

Figure~\ref{fig:sensitivity} and Table~\ref{tab:sens} vary each parameter around $K=5$, $m=1$, $h_0=3$, $\tau=0.35$, and Table~\ref{tab:removal} varies the support-removal margin $r$ of Appendix~\ref{app:removal}. CRNH counts admitted notes, shared or added, that X eventually rated \textsc{not helpful} (37 at the default, 31 of them additions). Fail at lock is the share of notes admitted before lock that no longer satisfy the rule on lock-time evidence (22.9\% of 99,738 at the default), which changes little with $K$ but falls with $m$.

\begin{table}[H]
\centering
\caption{Sensitivity and ablations. Bold is default/chosen values. Quality with 95\% intervals.}
\label{tab:sens}
\resizebox{\textwidth}{!}{%
\begin{tabular}{@{}llrrcrrr@{}}
\toprule
Param. & Value & Added (+\% over production) & +\% excl.\ firm-reject & Quality [95\% CI] & CRNH & Lead (h) & Fail at lock (\%) \\
\midrule
$K$ & 1 & 64,296 (+70.1\%) & +40.6\% & 0.845 [0.839, 0.851] & 83 & 1.72 & 23.6 \\
$K$ & 2 & 50,002 (+54.5\%) & +33.2\% & 0.851 [0.846, 0.857] & 54 & 1.35 & 22.8 \\
$K$ & 3 & 42,829 (+46.7\%) & +29.2\% & 0.855 [0.849, 0.861] & 47 & 1.13 & 22.6 \\
$K$ & \textbf{5} & \textbf{35,945 (+39.2\%)} & \textbf{+25.1\%} & \textbf{0.858 [0.852, 0.865]} & \textbf{37} & \textbf{0.88} & \textbf{22.9} \\
$K$ & 10 & 27,819 (+30.3\%) & +20.1\% & 0.859 [0.852, 0.867] & 32 & 0.57 & 23.2 \\
$K$ & 20 & 20,879 (+22.8\%) & +15.5\% & 0.863 [0.854, 0.871] & 22 & 0.31 & 23.7 \\
$K$ & 50 & 13,632 (+14.9\%) & +10.4\% & 0.862 [0.850, 0.875] & 16 & 0.01 & 25.0 \\
\midrule
$m$ & 0 & 139,019 (+151.7\%) & +70.5\% & 0.827 [0.820, 0.833] & 243 & 3.32 & 29.3 \\
$m$ & 2 & 8,588 (+9.4\%) & +6.1\% & 0.850 [0.837, 0.861] & 18 & $-0.51$ & 20.5 \\
\midrule
$h_0$ & 1 & 35,968 (+39.2\%) & +25.1\% & 0.858 [0.852, 0.865] & 37 & 0.88 & 22.9 \\
$h_0$ & 2 & 35,968 (+39.2\%) & +25.1\% & 0.858 [0.852, 0.865] & 37 & 0.88 & 22.9 \\
$h_0$ & 4 & 35,160 (+38.4\%) & +24.7\% & 0.858 [0.851, 0.865] & 37 & 0.83 & 22.8 \\
$h_0$ & 6 & 29,007 (+31.6\%) & +20.4\% & 0.854 [0.847, 0.861] & 32 & 0.36 & 22.5 \\
\midrule
$\tau$ & 0.2 & 36,248 (+39.5\%) & +25.2\% & 0.857 [0.850, 0.864] & 40 & 0.91 & 22.9 \\
$\tau$ & 0.5 & 38,486 (+42.0\%) & +26.8\% & 0.856 [0.850, 0.862] & 41 & 0.98 & 22.5 \\
$\tau$ & 0.75 & 48,318 (+52.7\%) & +33.2\% & 0.853 [0.846, 0.859] & 45 & 1.38 & 21.9 \\
\bottomrule
\end{tabular}}
\end{table}

\begin{table}[H]
\centering
\caption{Support-removal margin $r$ at $K=5$, $m=1$; bold: chosen setting.}
\label{tab:removal}
\small
\begin{tabular}{@{}lrcrrr@{}}
\toprule
$r$ & Added (+\% over production) & Quality [95\% CI] & CRNH & Lead (h) & Fail at lock (\%) \\
\midrule
\textbf{0} & \textbf{35,945 (+39.2\%)} & \textbf{0.858 [0.852, 0.865]} & \textbf{37} & \textbf{0.88} & \textbf{22.9} \\
1 & 19,927 (+21.7\%) & 0.855 [0.847, 0.863] & 26 & 0.17 & 21.6 \\
2 & 10,141 (+11.1\%) & 0.841 [0.829, 0.853] & 21 & $-0.37$ & 20.1 \\
\bottomrule
\end{tabular}
\end{table}

\section{Comparison methods}
\label{app:baselines}

Each Core alternative we look at loosens one hyperparameter of our reimplementation of Core, starting from production's setting: its intercept threshold $\theta$, lowered from 0.40 to 0.20, or its intercept penalty $\lambda_{\mathrm{int}}$, from 0.15 to 0.005. The reimplementation fits model~\eqref{eq:mf} to convergence in float64, with rater parameters from all non-population-sampled ratings through August 2026. It keeps Core's main gates (at least five ratings, $|f_n|<0.5$ and the net-helpful minimums), but not its tag, incorrect-rating and low-diligence filters or the population-sampled ratings production also used, so it approximates Core rather than reproducing it. At production's settings it passes 59.7\% of the notes Core published, and 88\% of what it passes was published by some production scorer (55\% by Core itself). The version behind Figures~\ref{fig:waiting} and~\ref{fig:mfspace}, fit on all ratings, passes 77.8\% of Core's CRH notes, and 67.3\% of what it passes is Core CRH. The Gaussian alternative scores a reimplementation of X's Gaussian scorer once, at lock, with its threshold lowered from 0.62 to 0.40. Regardless, these reimplementations are used for ablations only, and none enter the rule or its comparison with production, which use X's recorded decisions.

Table~\ref{tab:matched} compares, at matched coverage, the notes each rule adds that the other does not, since totals can rest on different notes. This is the case for the Gaussian, as only 5.5\% of our exclusive notes have population-sampled grades (635 of 11,498), against 28.8\% of the Gaussian's (3,368 of 11,680), so notes both add make up 77\% of our graded additions but 39\% of the Gaussian's. Our total is therefore higher (0.910 against 0.875) although the Gaussian's exclusive notes grade higher.

\begin{table}[H]
\centering
\caption{Matched-coverage comparison. Last two columns: helpful-share difference between the notes only we add and those only the alternative adds (95\% interval), and the overlap of the two sets of additions.}
\label{tab:matched}
\resizebox{\textwidth}{!}{%
\begin{tabular}{@{}llcccrc@{}}
\toprule
Alternative & Frame & Setting & Ours: added / quality & Alt.: added / quality & Ours-only $-$ alt.-only & Jaccard \\
\midrule
Lower intercept threshold & replay & $\theta=0.31$ & 35,945 / 0.858 & 37,401 / 0.838 & $+0.034$ [$+0.018$, $+0.050$] & 0.40 \\
Lower intercept threshold & lock time & $\theta=0.32$ & 22,351 / 0.910 & 23,102 / 0.851 & $+0.079$ [$+0.064$, $+0.092$] & 0.34 \\
Smaller $\lambda_{\mathrm{int}}$ & replay & $\lambda_{\mathrm{int}}=0.05$ & 35,945 / 0.858 & 36,671 / 0.813 & $+0.106$ [$+0.091$, $+0.122$] & 0.34 \\
Smaller $\lambda_{\mathrm{int}}$ & lock time & $\lambda_{\mathrm{int}}=0.05$ & 22,351 / 0.910 & 23,911 / 0.847 & $+0.097$ [$+0.082$, $+0.115$] & 0.30 \\
Gaussian threshold & lock time & $0.56$ & 22,351 / 0.910 & 22,533 / 0.875 & $-0.021$ [$-0.040$, $-0.002$] & 0.32 \\
\bottomrule
\end{tabular}}
\end{table}

\section{Coordinated manipulation}
\label{app:attack}

\begin{figure}[t]
\centering
\includegraphics[width=\textwidth]{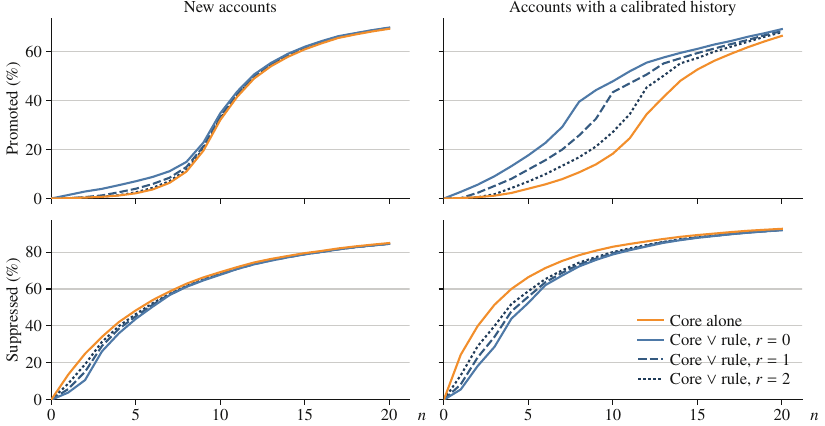}
\caption{Share of 10,000 targets flipped with at most $n$ coordinated accounts: pending notes promoted (top) and Core-published notes suppressed (bottom), for Core alone and for Core or the rule ($\lor$) with removal margin $r$.}
\label{fig:attack}
\end{figure}

We draw 10,000 targets of each kind with one seed shared across $r$: notes pending under Core and the base rule (promotion) and notes Core publishes (suppression). Core's response to the added ratings is simulated with our reimplementation (Appendix~\ref{app:baselines}). An adversary adds one rating from each of $n$ coordinated accounts, with calibration fixed as in (A1). Calibrated accounts carry the maximum capped weight in the rule but only 10th- or 90th-percentile parameters in Core, so the comparison is, if anything, pessimistic for the new rule. New accounts get the prior baseline and an attacker-chosen pole, a worst case beyond (A1): under our calibration they cannot move the rule, so the rule's new-account rows are hypothetical. Figure~\ref{fig:attack} and Table~\ref{tab:attack} give the results; a support-removal margin of $r=1$ cuts the additions to 19,927 and the median lead to 0.17 hours (Table~\ref{tab:removal}).

\begin{table}[H]
\centering
\caption{Targets flipped (\%) with at most $n$ coordinated accounts; the rule's new-account rows are a hypothetical worst case.}
\label{tab:attack}
\small
\begin{tabular}{@{}lllrrrrr@{}}
\toprule
Attack & Accounts & System & $n\le1$ & $n\le2$ & $n\le5$ & $n\le10$ & $n\le20$ \\
\midrule
Promotion & new & Core alone & 0.1 & 0.4 & 2.1 & 32.1 & 69.5 \\
 & & Core $\lor$ rule, $r=0$ & 1.4 & 2.9 & 7.1 & 34.9 & 70.0 \\
 & & Core $\lor$ rule, $r=1$ & 0.1 & 0.6 & 4.0 & 33.6 & 69.8 \\
 & & Core $\lor$ rule, $r=2$ & 0.1 & 0.4 & 2.5 & 32.8 & 69.7 \\
 & calibrated & Core alone & 0.2 & 0.5 & 4.0 & 18.3 & 66.5 \\
 & & Core $\lor$ rule, $r=0$ & 2.7 & 5.7 & 17.7 & 48.0 & 69.3 \\
 & & Core $\lor$ rule, $r=1$ & 0.2 & 2.4 & 11.9 & 43.5 & 68.5 \\
 & & Core $\lor$ rule, $r=2$ & 0.2 & 0.5 & 7.0 & 27.2 & 68.1 \\
\midrule
Suppression & new & Core alone & 13.6 & 24.9 & 48.4 & 69.1 & 84.9 \\
 & & Core $\lor$ rule, $r=0$ & 3.9 & 10.7 & 43.8 & 67.7 & 84.4 \\
 & & Core $\lor$ rule, $r=1$ & 5.8 & 15.2 & 45.2 & 68.0 & 84.5 \\
 & & Core $\lor$ rule, $r=2$ & 8.7 & 19.2 & 46.4 & 68.2 & 84.5 \\
 & calibrated & Core alone & 24.4 & 39.9 & 66.4 & 82.9 & 92.6 \\
 & & Core $\lor$ rule, $r=0$ & 5.6 & 18.3 & 52.5 & 78.7 & 91.8 \\
 & & Core $\lor$ rule, $r=1$ & 8.3 & 22.9 & 55.9 & 79.5 & 91.9 \\
 & & Core $\lor$ rule, $r=2$ & 13.3 & 29.0 & 58.8 & 80.0 & 92.0 \\
\bottomrule
\end{tabular}
\end{table}


\end{document}